\documentclass[lettersize,journal]{IEEEtran}
\usepackage{amsmath,amsfonts,amssymb}
\usepackage{algorithmic}
\usepackage{algorithm}
\usepackage{array}
\usepackage[caption=false,font=normalsize,labelfont=sf,textfont=sf]{subfig}
\usepackage{textcomp}
\usepackage{stfloats}
\usepackage{url}
\usepackage{verbatim}
\usepackage{graphicx}
\usepackage{cite}
\usepackage[dvipsnames]{xcolor}
\newcommand{\ignore}[1]{}
\newcommand{\fixme}[1]{\textcolor{red}{#1}}
\begin{document}

\title{A 0.65-pJ/bit 3.6-TB/s/mm I/O Interface with XTalk Minimizing Affine Signaling for Next-Generation HBM with High Interconnect Density}

\author{Hyunjun Park,~\IEEEmembership{Graduate Student Member,~IEEE}, Jiwon~Shin,~\IEEEmembership{Member,~IEEE},

Hanseok~Kim,~\IEEEmembership{Graduate~Student~Member,~IEEE}, Jihee~Kim,~\IEEEmembership{Graduate Student Member,~IEEE}, 

Haengbeom Shin,~\IEEEmembership{Graduate Student Member,~IEEE}, Taehoon Kim,~\IEEEmembership{Graduate Student Member,~IEEE}, 

Jung-Hun Park,~\IEEEmembership{Member, IEEE},  and~Woo-Seok~Choi,~\IEEEmembership{Member,~IEEE}
\vspace{-2em}
\thanks{This work was supported by the Samsung Electronics Company, Ltd., Hwaseong, Korea.}
\thanks{H. Park, J.-H. Park, J. Shin, H. Kim, J. Kim, H. Shin, T. Kim, and W.-S. Choi are with the Department of Electrical and Computer Engineering and the Inter-University Semiconductor Research Center, Seoul National University, Seoul 08826, South Korea (e-mail: spp098@snu.ac.kr, wooseokchoi@snu.ac.kr).}}

\markboth{Journal of \LaTeX\ Class Files,~Vol.~14, No.~8, August~2021}
{Shell \MakeLowercase{\textit{et al.}}: A Sample Article Using IEEEtran.cls for IEEE Journals}

\maketitle

\begin{abstract}
This paper presents an I/O interface with Xtalk Minimizing Affine Signaling (XMAS), which is designed to support high-speed data transmission in die-to-die communication over silicon interposers or similar high-density interconnects susceptible to crosstalk. 
The operating principles of XMAS are elucidated through rigorous analyses, and its advantages over existing signaling are validated through numerical experiments. 
XMAS not only demonstrates exceptional crosstalk removing capabilities but also exhibits robustness against noise, especially simultaneous switching noise. 
Fabricated in a 28-nm CMOS process, the prototype XMAS transceiver achieves an edge density of 3.6\,TB/s/mm and an energy efficiency of 0.65\,pJ/b. 
Compared to the single-ended signaling, the crosstalk-induced peak-to-peak jitter of the received eye with XMAS is reduced by 75\,\% at 10\,GS/s/pin data rate, and the horizontal eye opening extends to 0.2\,UI at a bit error rate $<$ 10\text{$^{-12}$}.

\ignore{This paper introduces the Xtalk Minimizing Affine Signaling (XMAS) transceiver, which is designed to support high-speed transmission in die-to-die communication over silicon interposers or similar high-density interconnects while effectively mitigating crosstalk. The operating principles of XMAS are elucidated through rigorous analyses, and its advantages are substantiated through numerical experiments. XMAS not only demonstrates exceptional crosstalk removal capabilities but also exhibits robustness against noise, particularly owing to its permutability, which makes it highly resilient to simultaneous switching noise. Fabricated using a 28-nm CMOS process, the XMAS transceiver achieves a shoreline throughput of 3.6 TB/s/mm and boasts an energy efficiency of 0.65 pJ/b. The peak-to-peak jitter of the received eye is reduced by 75\% at a 10 GS/s data rate, and the horizontal eye opening extends to 0.2 UI at a bit error rate (BER) of less than 10\text{$^{-12}$}.}
\end{abstract}

\section{Introduction}
\IEEEPARstart{H}{igh}-performance computing (HPC) catalyzes transformative developments across diverse domains, encompassing artificial intelligence, cloud computing, natural sciences such as astronomy and physics, and humanities including economics and sociology. In light of its significance, there is a pressing need to enhance HPC's capabilities and address its technological challenges. Nevertheless, as the CMOS technology scaling slows down, the effort for enhancement faces obstacles. Acknowledging these challenges, the emergence of advanced packaging technologies offers promising avenues to sustain technological progress and prolong Moore's Law~\cite{mirabbasi2022through}.

High Bandwidth Memory (HBM), where data between a host and memory are transmitted over thousands of silicon interposer channels, can offer high bandwidth suitable for HPC applications,
but the demands for even higher bandwidth are rapidly growing to support emerging applications.
To maintain overall power budget, higher bandwidth demand should be accompanied by I/O energy efficiency scaling in HBM. 
Pursuing high bandwidth with low area and energy consumption of the I/O interface leads to the adoption of single-ended (SE) signaling, which exhibits 2$\times$ higher pin efficiency than differential signaling for data transmission, over unterminated channels~\cite{8260536,7418035}.
However, SE brings several disadvantages, particularly its vulnerability to various noise. 
In systems like HBM having extremely large number of I/O's, the signal deterioration due to data-dependent simultaneous switching noise (SSN) becomes especially pronounced. 
Various strategies have been explored and developed to mitigate these challenges~\cite{poulton20130,poulton20181,kwon202333,shokrollahi201610,tajalli20191,tajalli2020short,tajalli20201,cronie2011orthogonal}.

Another key obstacle to higher bandwidth is the crosstalk (XTalk) between the channels. 
Higher edge density requires small spacing between the channels, increasing crosstalk between neighboring channels~\cite{lee20135,rainal1979transmission}, 
which becomes a major threat to signal integrity. 
To meet the ever-growing demand for higher bandwidth, we should maximize throughput and ensure robustness against noise and crosstalk, while not sacrificing pin efficiency.
Addressing this pivotal question forms the core objective of this paper, introducing XTalk Minimizing Affine Signaling (XMAS) as a novel solution to this multifaceted challenge.

Our approach begins with a comprehensive mathematical modeling of XMAS to capture the system performance such as eye width and eye height at the receiver in the presence of crosstalk. 
This modeling allows design space exploration and strategic co-optimization of the channel and XMAS design 
to achieve the highest edge density without compromising signaling integrity. 

Compared to prior art employing coding or circuit-level techniques for crosstalk cancellation (XTC),
the proposed XMAS shows better performance as follows.
Conventional bus encoding techniques~\cite{liu20230,duan2001analysis,victor2001bus,subrahmanya2004bus} add redundant bits to reduce crosstalk, which significantly compromises pin efficiency. 
Furthermore, as it inherently adopts SE signaling, it is susceptible to noise, making it less robust in systems like HBM. 
Another prevalent approach involves direct compensation of distortion caused by crosstalk using equalizers~\cite{kao20127,ko20206,aprile2018eight,oh201312,oh20116,nazari201115gb,kao201310,zhong20222}. 
Despite its effectiveness, this method introduces significant hardware complexity and overhead, worsening the overall I/O energy efficiency.
Unlike these methods, XMAS assigns optimized correlation across multiple wires, achieving a remarkable pin efficiency of 87.5\,\%. 
Moreover, XMAS ensures robustness against noise and crosstalk without incurring circuit overhead. 
This novel approach not only addresses the limitations of the previous methods but also offers a balance between pin efficiency and robustness against noise and crosstalk. 

The remainder of this paper is organized as follows. Section~\ref{sec:signaling} presents a mathematical model for XMAS, which serves as the foundation for the co-optimization techniques with the channel detailed in Section~\ref{sec:optimization}. Section~\ref{sec:implementation} introduces the XMAS transmitter and receiver implementations, and Section~\ref{sec:measured} presents the measurement results of the prototype transceiver, followed by a conclusion of this work in Section~\ref{sec:conclusion}.

\section{Mathematical Modeling}
\label{sec:signaling}

\begin{figure}[t!]
\centering
\includegraphics[width=\columnwidth]{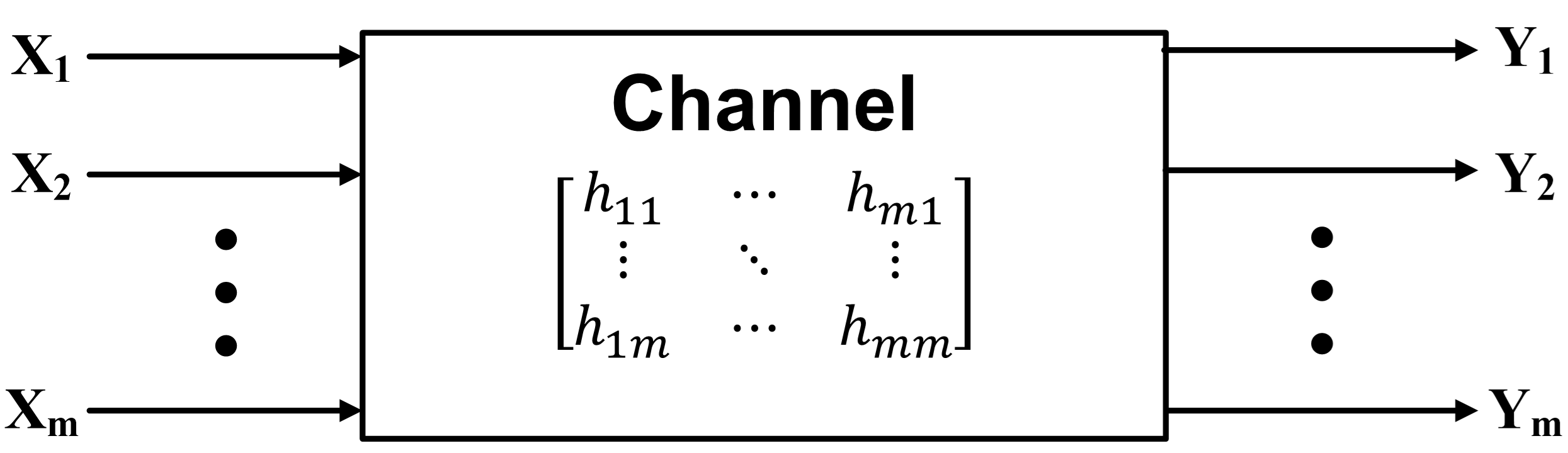}
\caption{Multi-input multi-output linear time invariant channel.}
\label{fig:mimo}
\end{figure}

In XMAS, the transmitter (TX) transmits voltage levels after applying an affine transformation to the incoming parallel binary data, and the receiver (RX) recovers the binary data by linearly transforming the received voltage levels. In the following, we present an analytical model for XMAS that enables cooptimization of signaling and interconnect design presented in Section~\ref{sec:optimization}.
\ignore{In the following, we present an analytical model for XMAS 
by which we can accurately calculate the impact of crosstalk on signal integrity and suggest how to find optimal transformations (or matrices) for encoding and decoding that can minimize the crosstalk. }


\subsection{Channel Modeling}

Parallel channels with crosstalk can be considered a multi-input multi-output (MIMO) linear time-invariant (LTI) system, which can be described using a set of channel impulse response as shown in Fig.~\ref{fig:mimo}. 
($h_{ij}$ denotes the channel output $Y_j$ when the input $X_i$ is given as an impulse.)
Specifically, when the channel receives the pulse-amplitude-modulated signals, i.e., $X_{l}(t) = \sum_{i=-\infty}^{\infty} a_{l,i} \Pi(t-iT)$ where $\Pi(t)$ is 1 for $0\leq t < T$ and 0 otherwise, the output can be represented as follows. 
\begin{align}
Y_{j} &= \sum_{i=1}^{m} X_{i}(t)*h_{ij}(t)=\sum_{i=1}^{m}\sum_{k=-\infty}^{\infty}a_{i,k}E_{ij}(t-kT).
\label{eq:channel}
\end{align}
$E_{ij}(t)$ denotes the response of the $j$-th channel to a single-bit-pulse input originating from the $i$-th channel. 
If the channel loss is sufficiently small (i.e., intersymbol interference is negligible), the $k$-th output of $j$-th channel depends only on the $k$-th $m$-parallel inputs $a_{i,k} (i\in [m])$ and $E_{ij}(t)$. 
For a MIMO channel,
we define $\mathbf{H_k}$ as an $m \times m$ matrix whose elements are $E_{ij}(t-kT)$.

\ignore{
\begin{align}    
& \mathbf{H_{i}}=[E_{kl}(t-iT)]_{m\times m},~\mathbf{S_i}=diag(\mathbf{H_i})\notag\\
& \forall~k\neq l,~E_{kl}(t)=c_{kl}(t)-c_{kl}(t-T)\notag\\
&\mathbf{C_{i}}-\mathbf{C_{i+1}}=\mathbf{H_i}-\mathbf{S_i}\notag\\
&\mathbf{C_{i}}=[\alpha_{kl}],~\alpha_{kl}=
\begin{cases}
   0& \text{if } k=l\\
   c_{kl}              & \text{otherwise}
\end{cases}\notag\\
&length~of~supp~E_{kl}(t-iT)\leq 2T
\end{align}
}

\subsection{Encoding/Decoding with Affine/Linear Transformation}
\begin{figure}[!t]
\centering
\includegraphics[width=\columnwidth]{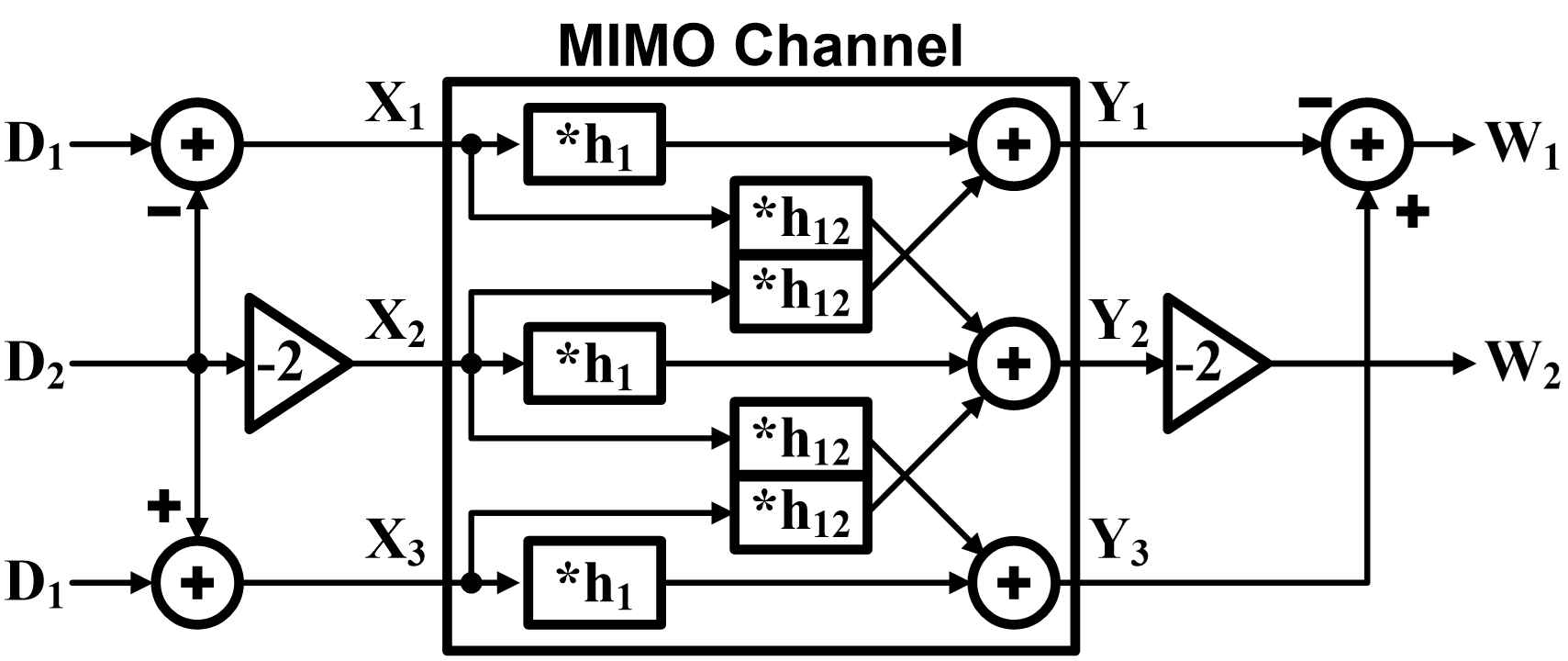}
\caption{XMAS example $(n=3, m=2)$.}
\label{fig:XMASex}
\end{figure}

\begin{figure}[!t]
\centering
\includegraphics[width=\columnwidth]{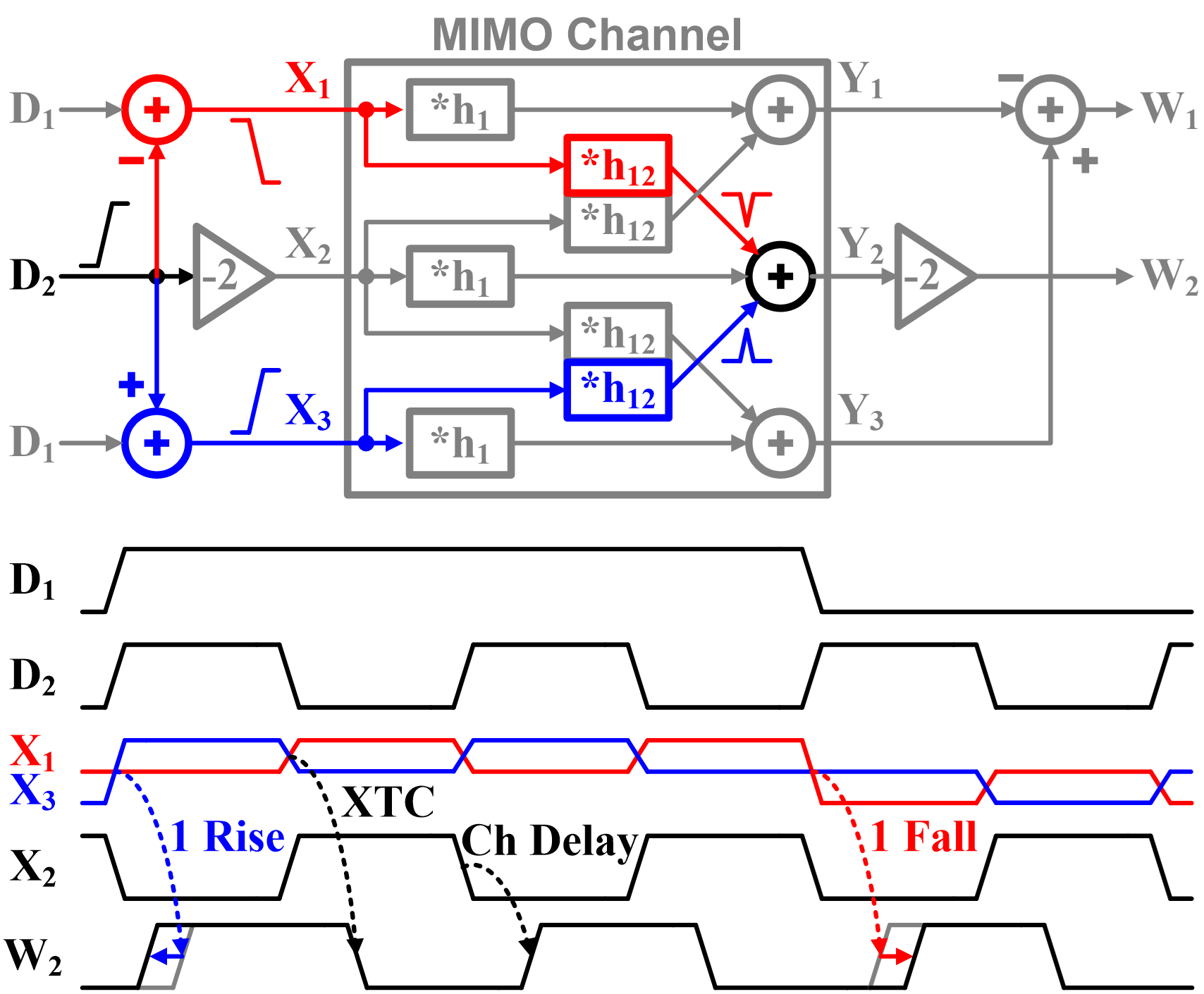}
\caption{XTC for $W_2$.}
\label{fig:xtcprin1}
\end{figure}
\begin{figure}[!t]
\centering
\includegraphics[width=\columnwidth]{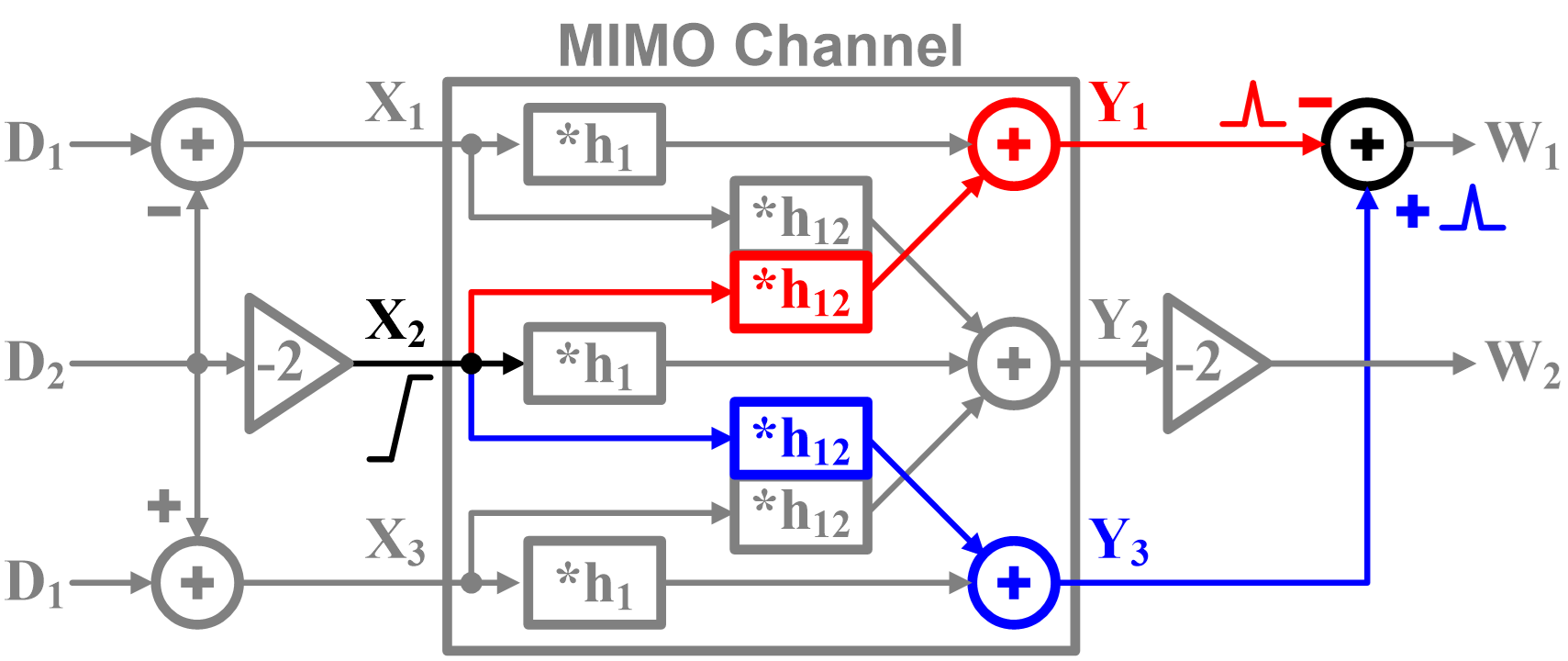}
\caption{XTC for $W_1$.}
\label{fig:xtcprin2}
\end{figure}
In XMAS, an affine transformation is applied to the parallel input data, which are then transmitted by TX using pulse-amplitude modulation.
Specifically, an $n \times m$ integer matrix $\mathbf{T}$ is used to encode $m$-parallel incoming binary data to an $n$-dimensional integer vector.
Then, the encoded elements are mapped to the voltage levels $\mathbf{a_{i}}$ between 0 and $V_{DDQ}$ (supply voltage of the TX output driver),
which can be expressed as:
\begin{align}
\mathbf{a_{i}}&=
\left[
\begin{matrix}
a_{1,i}  
\hdots 
a_{n,i} 
\end{matrix}
\right]^\intercal =  0.5\cdot V_{DDQ}(\mathbf{T_{eff}}\mathbf{d_{i}}+\left[
\begin{matrix}
1  
\hdots 
1 
\end{matrix}
\right]^\intercal)
\label{eq:affine}
\end{align}
where $\mathbf{d_i} \in \{-1, 1\}^m$ denotes a vector representing $m$-parallel binary input to be transmitted, 
and $\mathbf{T_{eff}}$ is the normalized $\mathbf{T}$ such that the $\ell$1-norm of each row vector in $\mathbf{T_{eff}}$ becomes 1 to gurarantee that the voltage levels $\mathbf{a_{i}}$ are within 0 and $V_{DDQ}$. \ignore{For instance, if two inputs $d_{i,1}$ and $d_{i,2}$ are transformed by a row vector [1 -1],
then the output voltage becomes $0.25V_{DDQ}(d_{i,1} -d_{i,2}) + 0.5V_{DDQ}$.
\eqref{eq:affine} guarantees that an encoded element can be mapped to a voltage between 0 and $V_{DDQ}$. 
TX transmits the pulse-amplitude-modulated signal with $\mathbf{a_i}$ to the $n$ channels.}\ignore{
\begin{align}
\boldsymbol{D_{i}} &=[d_{j,i}]_{n\times 1, j=1,\cdots,n},~d_{j,i} \in \{-1,1\}\notag\\
\mathbf{T_{eff}} &=diag(t_{1}^{-1},t_{2}^{-1},\dots,t_{n}^{-1})\mathbf{T}\notag\\
t_{i}&=\left[
\begin{matrix}
\delta_{i1}\;\delta_{i2}\;\cdots\;\delta_{in} \\
\end{matrix}
\right]abs(\mathbf{T})\left[
\begin{matrix}
1  \\
\vdots \\
1 \\
\end{matrix}
\right]
\end{align}
}
Then the channel outputs can be represented as $\mathbf{H_{i}a_{i}}$,
which undergoes a linear transformation at the RX front end.
If the linear transformation applied by RX is an $m \times n$ integer matrix $\mathbf{R}$,
then the decoded outputs are\footnote{In \eqref{eq:output}, the bias term due to $[1 \hdots 1]^\intercal$ in \eqref{eq:affine} is omitted since $\mathbf{R}$ will be chosen to make the bias term become zero.}
\begin{align}
\mathbf{W}=\sum_{i=-\infty}^{\infty}\mathbf{RH_{i}a_{i}} &= \sum_{i=-\infty}^{\infty}0.5V_{DDQ}(\mathbf{RH_{i}T_{eff}d_i}).
\label{eq:output}
\end{align}

As a toy example, Fig.~\ref{fig:XMASex} illustrates XMAS with below matrices, where two input data are encoded over three channels. 
\begin{align}
&\mathbf{T}=\begin{bmatrix}
1 & -1 \\
0 & -2 \\
1 & 1 \\
\end{bmatrix} \notag
~\mathbf{R}=\begin{bmatrix}
-1 & 0 & 1 \\
0 & -2 & 0 \\
\end{bmatrix}\notag \\ 
&\mathbf{H}=\begin{bmatrix}
h_1 & h_{12} & 0 \\
h_{12} & h_1 & h_{12} \\
0 & h_{12} & h_1 \notag
\end{bmatrix}
\notag
\end{align}
Inputs $D_1$ and $D_2$ are encoded by $\mathbf{T}$ into a set of signals $(X_1, X_2, X_3)$ transmitted through the channels. 
The channel characteristics are defined by $\mathbf{H}$, where each channel has identical pulse response $h_1$, and the symmetric coupling between adjacent channels is denoted by $h_{12}$. 
Channel outputs $(Y_1, Y_2, Y_3)$ are then decoded by matrix $\mathbf{R}$ into symbols $W_1=\hat{D_1}$ and $W_2=\hat{D_2}$. 
Due to the channel structure, the channel inputs $X_1$ and $X_3$ influence $Y_2$, causing crosstalk-induced jitter (CIJ). 
However, as depicted in Fig.~\ref{fig:xtcprin1}, the input $D_2$ is added with the opposite signs into the red and blue paths, effectively canceling the crosstalk at $Y_2$. 
In more detail, when input data in Fig.~\ref{fig:xtcprin1} are given, $X_1$ and $X_3$ transition in the exactly opposite directions, perfectly canceling out the crosstalk, or when one signal $(X_1)$ undergoes a full swing transition, the transition in the other signal $(X_3)$ is always prevented, thereby reducing CIJ. 
Another source of crosstalk is the influence of $X_2$ on $Y_1$ and $Y_3$. 
As shown in Fig.~\ref{fig:xtcprin2}, $X_2$ causes the same amount of distortion in $Y_1$ and $Y_3$, which is perfectly canceled during decoding. 
Such properly designed XMAS matrices thus can hold significant potential for XTC. 
Hence, in the following, we show how to carefully design the XMAS matrices with the channels to maximize the interface edge density by taking advantage of excellent XTC with XMAS.

\section{XMAS Design}
\label{sec:optimization}

\begin{figure*}[!t]
\centering
\includegraphics[width=\textwidth]{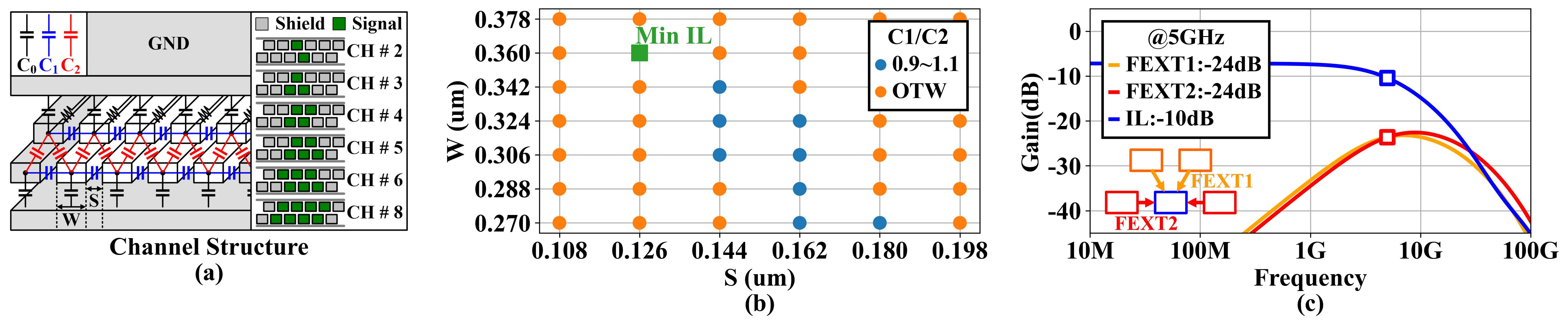}
\caption{Channel design parameter $(S, W)$ optimization: (a) Physical structure of channels. (b) Coupling capacitance ratio ($C_1/C_2$) between adjacent channels as channel width ($W$) and spacing ($S$) vary. (c) Channel transfer function with $(S,W,L)$ = (0.126\,{\textmu}m, 0.36\,{\textmu}m, 1.26\,mm).}
\label{fig:channelopt}
\end{figure*}
\begin{figure*}[!t]
\centering
\includegraphics[width=\textwidth]{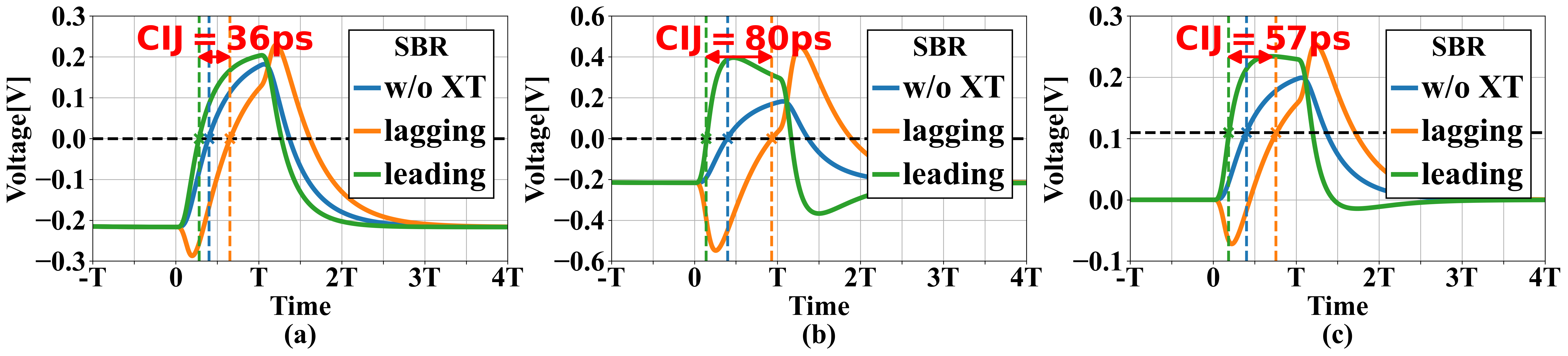}
\caption{Impact of crosstalk-induced jitter on SBR: (a) XMAS with optimized matrices. (b) XMAS with degenerate matrices. (c) Single-ended signaling.}
\label{fig:cij}
\end{figure*}

This section focuses on determining the XMAS and channel design parameters, aiming to maximize edge density in dense channel environments. 
Edge density is affected by various design parameters such as channel dimensions, per-pin data rate, and XMAS matrices. 
The intricate interplay of diverse parameters significantly influences edge density, making it complex and challenging to find optimal parameters achieving the best performance. 
Moreover, since the number of possible encoding and decoding matrices in XMAS is vast and 
simulating each case is computationally complex and time-consuming, 
finding an optimal parameter set without a theoretical model is practically infeasible. 
Thus, the analytical model described in Section~\ref{sec:signaling} is leveraged to efficiently navigate the expansive parameter space and to find the optimal parameters for the maximum edge density. 
Specifically, the optimization problem for the prototype XMAS transceiver is defined as follows:
\begin{align}
    \max_{S, W, L, \mathbf{T_{n\times m}, R_{m\times n}}, B} \quad & \text{Edge Density} \label{eq:optprob}\\
    \text{subject to} \quad & \text{Eye Width} \geq 0.7 \, \text{UI, Height} \geq 100 \, \text{mV}\notag \\
    & \text{Channel Loss} \leq 10 \, \text{dB, } 0.9 \leq \frac{C_1}{C_2} \leq 1.1 \notag 
\end{align}
where $S$, $W$, $L$, $C_1$, and $C_2$ denote the channel spacing, width, length, capacitance between the adjacent channels in the same layer and different layers, respectively (see Fig.~\ref{fig:channelopt}(a)), and $B$ represents the symbol rate (Baud). 
Note that the last two constraints in \eqref{eq:optprob} are added so the designed channels have low loss and symmetric capacitance between adjacent channels. 

\subsection{Codesign of Interconnects with Signaling}

Fig.~\ref{fig:channelopt}(a) depicts the adopted channel layout, following the densely structured approach of \cite{ko20206} to maximize the interconnect density. 
The number of wires grouped for encoded data transmission, determined by the number of rows ($n$) in the XMAS encoding matrix $\mathbf{T}$, can be adjusted to alter the channel configuration, and the width ($W$) and spacing ($S$) of the channels determine their characteristics such as the channel resistance and capacitance as well as the coupling capacitance between adjacent channels. 
Fig.~\ref{fig:channelopt}(b) illustrates the variation of the coupling capacitance ratio $C_1/C_2$ as $W$ and $S$ change. 
A ratio close to 1 indicates symmetric crosstalk from adjacent channels, represented as the blue dots in Fig.~\ref{fig:channelopt}(b)\footnote{Although not mandatory, symmetric crosstalk yields simpler XMAS design, so the ratio close to 1 is chosen for the prototype implementation.}. 
Among them, the configuration with the lowest insertion loss (IL), which corresponds to $S$ = 0.126\,{\textmu}m and $W$ = 0.36\,{\textmu}m, is marked with the green square and chosen for the prototype. 
When the eight wires are designed with $(S,W,L)$ = (0.126\,{\textmu}m, 0.36\,{\textmu}m, 1.26\,mm), the channel characteristics are obtained as Fig.~\ref{fig:channelopt}(c). 
At a frequency of 5\,GHz, the insertion loss is around 10\,dB, and the far-end crosstalk (FEXT) reaches $-24$\,dB, primarily emanating from the four adjacent channels. 
These channels emerge as the principal sources of interference, whereas more distant channels have a negligible impact on the overall crosstalk.

\begin{figure*}[!t]
\centering
\includegraphics[width=\textwidth]{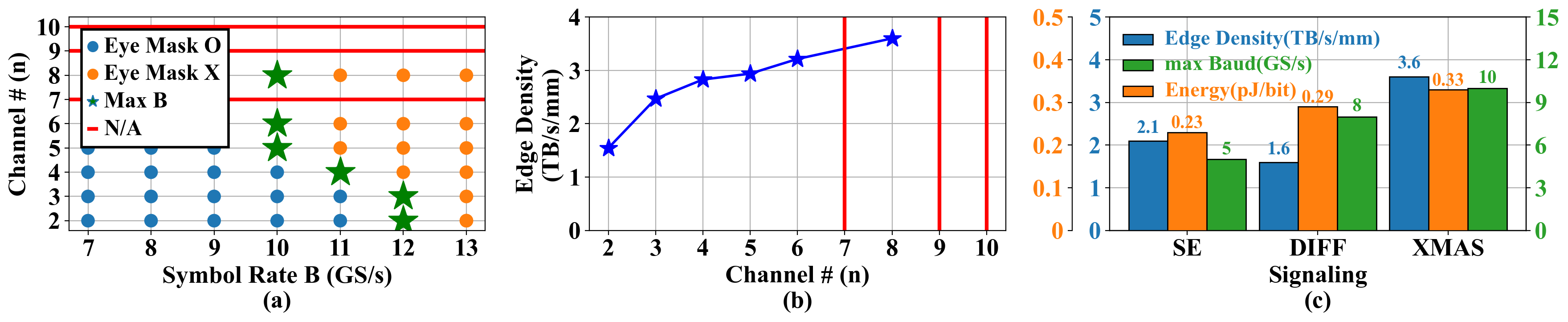}
\caption{XMAS parameter optimization $(n,m,B)$: (a) Representing whether required eye mask is satisfied or not as symbol rate ($B$) and number of channels ($n$) vary. (b) Maximum achievable edge density for different number of channels. (c) Comparison of edge density, maximum symbol rate, and energy efficiency between different signaling schemes. }
\label{fig:optimization}
\end{figure*}

\subsection{XMAS Matrix Design}

\begin{figure}[!t]
\centering
\includegraphics[width=\columnwidth]{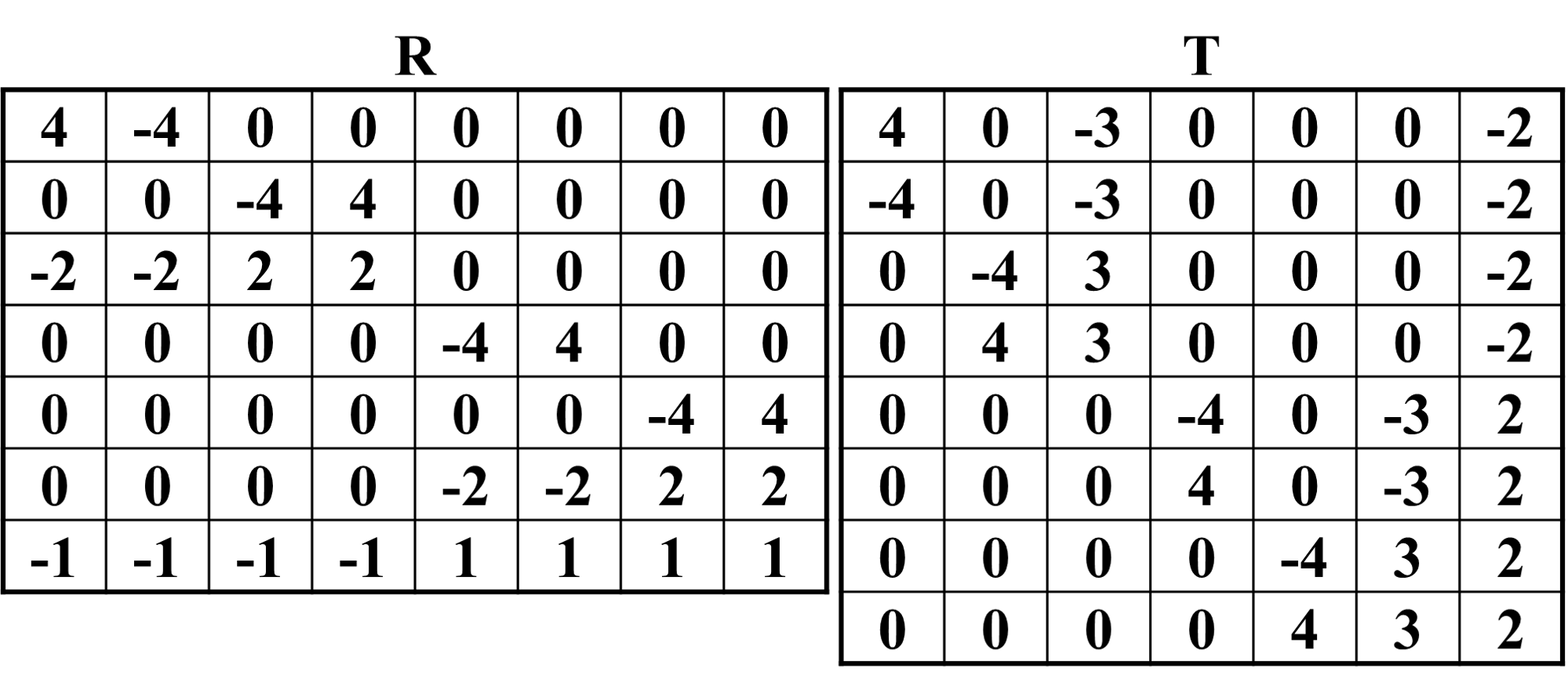}
\vspace{-2em}
\caption{Proposed XMAS matrices.}
\label{fig:matrix}
\end{figure}

For the described channel configuration, the XMAS matrices $\mathbf{T}$ and $\mathbf{R}$ are designed to possess the following properties that significantly improve signal integrity. 

\subsubsection{Binary Decision} 
Although pulse-amplitude-modulated signals are transmitted through channels, 
RX makes a binary decision with XMAS, which minimizes the sensitivity to intersymbol interference due to channel loss, similar to chord signaling~\cite{tajalli20201}.
To this end, the XMAS encoding and decoding matrices, $\mathbf{T}$ and $\mathbf{R}$, are designed such that every row vector in $\mathbf{R}$ and every column vector in $\mathbf{T}$ are orthogonal. 
In other words, for some diagonal matrix $\mathbf{\Lambda}$,
\begin{equation}
    \mathbf{R}\mathbf{T} = \mathbf{\Lambda}.    
    \label{eq:orthogonal}
\end{equation}
With this condition, due to \eqref{eq:output}, the decoded outputs will have binary levels. 

\subsubsection{Minimal Crosstalk-Induced Jitter (CIJ)} 

\ignore{
\begin{align}
0.5V_{DDQ}\sum_{i=-\infty}^{\infty}\boldsymbol{DBR_{i}}& =[b_{j}]_{n\times1,j=1,2,\cdots,n}\notag\\
x_{ij}& =b_{j}|_{d_{i,j}=1, d_{\neq i,j}=-1}\notag\\
y_{ij}& =b_{j}|_{d_{i,j}=-1, d_{\neq i,j}=1}
\end{align}
}
For the designed channels, Fig.~\ref{fig:cij} shows the simulation result demonstrating the impact of CIJ on single-bit response (SBR), where the zero-crossing times are either delayed or advanced depending on the data pattern. 
This distorted SBR can be accurately captured by substituting appropriate patterns into $\mathbf{d_i}$ as each element of $\mathbf{W}$ in \eqref{eq:output} represents the output waveform at the RX. 
\ignore{For instance, we can get the worst-case zero-crossing time on data $d_{4}$ by calculating \eqref{eq:output} after setting $\mathbf{d}$ such that $d_{4} = 1$ and $d_{j} = -1$ for $j \neq 4$.
Similarly, the best-case zero-crossing time can be computed by setting all the elements in $\mathbf{d}$ as 1.
Since CIJ is the difference between the best and the worst zero-crossing times, CIJ for each data can be obtained using \eqref{eq:output}. Note from \eqref{eq:output} that CIJ, or the zero-crossing timing difference ($J_j$), is determined by the encoding and decoding matrices, $\mathbf{T}$ and $\mathbf{R}$, and the MIMO channel SBR matrix $\mathbf{H}$.
In other words, we need to choose optimal $\mathbf{T}$ and $\mathbf{R}$ that minimize CIJ, or $J_j$, for given channels.
As illustrated in Fig.~\ref{fig:cij}, with optimal selection of $\mathbf{T}$ and $\mathbf{R}$, CIJ can be greatly reduced; 
Otherwise, CIJ will become worse than SE.}
For instance, CIJ-induced SBR for the channel \#4 ($W_4$) can be calculated by evaluating \eqref{eq:output} after setting $\mathbf{d_{i}}$ such that a single bit pulse is given to $d_4$ (i.e., $d_{i,4} = 1$ and $d_{j,4} = -1$ for $j \neq i$), and all the possible combinations of data patterns are provided to the other data.
Then, $\textrm{CIJ}_4$, which is the largest difference between the zero-crossing times for $W_4$, can be readily calculated based on the computed SBRs. 
Note that, since the SBR waveforms are determined by $\mathbf{T}$ and $\mathbf{R}$, 
proper matrix selection for given channels ($\mathbf{H_i}$) helps minimize CIJ.
As illustrated in Fig.~\ref{fig:cij}, with the optimal selection of $\mathbf{T}$ and $\mathbf{R}$, CIJ can be greatly reduced; otherwise, CIJ can become worse than SE.

\ignore{
\begin{align}
\color{blue} x_{ij} &= \color{blue}0.5V_{DDQ}\mathbf{R_{eff}(H_{j}-\sum_{k\neq j}H_{k})T_{eff}d_j}|_{d_{j,i}=1}\notag
\label{eq:output2}
\end{align}
}


\subsubsection{Minimal SSN} 
The encoding matrix in XMAS is constructed so the set of voltage levels formed by the $n$ TX drivers can be always constant, 
i.e., each driver may transmit different voltage during each unit interval (UI), 
but as a group of $n$ drivers, they always transmit the identical voltages. 
For instance, for the $7\times8$ encoding matrix $\mathbf{T}$ presented in Fig.~\ref{fig:matrix}, 
eight voltages of $V_{DDQ}\cdot[0, 2/9, 3/9, 4/9, 5/9, 6/9, 7/9, 1]$ are invariably used by eight drivers. 
In other words, regardless of seven input data, the eight encoded data have one of the values in $V_{DDQ}\cdot[0, 2/9, 3/9, 4/9, 5/9, 6/9, 7/9, 1]$, and each driver transmits distinct voltage levels.
This ensures that the current provided from the supply to the drivers remains constant irrespective of input data, thereby greatly removing SSN. 

\ignore{In this section, a rigorous mathematical model of XMAS has been developed, enabling accurate calculation of the signal integrity of XMAS. The results show that it is feasible to determine the matrices T and E that yield the least crosstalk-induced jitter by calculating every possible transformation. This meticulous approach enhances the preciseness of the signal integrity assessment and fosters a robust foundation for optimizing the encoding and decoding matrices. }

Having discussed the desired XMAS properties, now we describe how to design $\mathbf{T}$ and $\mathbf{R}$ for XMAS. 
Once the channel dimensions $S, W$, and $L$ are determined, for the fixed values of $m$ and $n$, integer matrices $\mathbf{T}$ and $\mathbf{R}$ can be determined to satisfy the required properties. 
To improve pin efficiency and maximize edge density, the number of parallel data to be encoded ($m$) is fixed at $n-1$. 
Fig.~\ref{fig:optimization}(a) depicts the symbol rates ($B$) that satisfy the eye mask constraints in \eqref{eq:optprob}, with each set of optimized matrices $\mathbf{T}$ and $\mathbf{R}$ uniquely determined for the fixed channel configuration across various values of $n$. 
Note that, for $n$ equal to 7 or greater than 8, the orthogonality condition \eqref{eq:orthogonal} is unattainable, which is illustrated by a red line in Fig.~\ref{fig:optimization}(a). 
As $n$ decreases, the maximum symbol rate meeting the eye mask constraint tends to increase, suggesting that a smaller $n$ yields better XTC performance of XMAS. 
However, this also results in lower pin efficiency, thereby not necessarily yielding the highest edge density. 
Fig.~\ref{fig:optimization}(b) illustrates the edge density at the maximum symbol rate for each $n$ value. 
While the XTC performance of XMAS may decrease as $n$ increases, higher pin efficiency leads to an increase in edge density. 
Consequently, for maximum edge density, we choose $n$ = 8 and the corresponding optimal orthogonal matrices $\mathbf{T}$ and $\mathbf{R}$ are depicted in Fig.~\ref{fig:matrix}. 
For the identical channel configuration and eye-opening conditions, XMAS outperforms SE and differential signaling thanks to its high XTC performance even at a maximum symbol rate of 10\,GS/s. 
Moreover, its high pin efficiency (7/8) allows XMAS to achieve 1.7× and 2.25× higher edge density compared with SE and differential signaling, respectively, while maintaining comparable energy efficiency, as illustrated in Fig.~\ref{fig:optimization}(c).

\section{XMAS Interface Architecture}
\label{sec:implementation}
\begin{figure*}[!t]
\centering
\includegraphics[width=\textwidth]{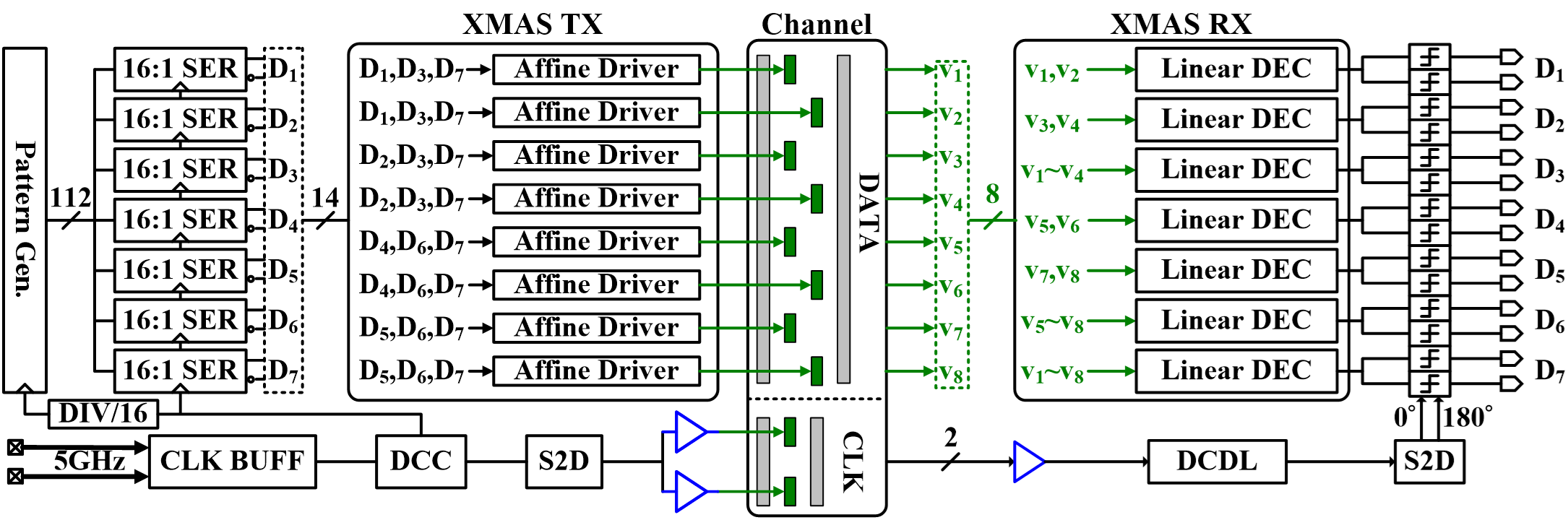}
\caption{Overall architecture of the proposed transceiver with XMAS.}
\label{fig:arch}
\end{figure*}

Fig.~\ref{fig:arch} illustrates the overall architecture of the proposed interface with XMAS. 
TX serializes the $16\times7$ bits of parallel data by 16:1 serializers (SERs) and supplies 7 parallel differential data stream to the 8 affine drivers, which drive the channels with appropriate voltages. 
The affine drivers encode the parallel incoming data following the row vectors of $\mathbf{T}$ in Fig.~\ref{fig:matrix}. 
Since each row of $\mathbf{T}$ consists of three non-zero elements, 
each driver takes three out of seven parallel incoming data as shown in Fig.~\ref{fig:arch}.
At the RX front end, the linear decoders, implemented following the row vectors of $\mathbf{R}$, convert the received voltages from 8 channels into 7 binary symbols to restore the data. 
For the clock path, a limited-swing half-rate differential clock is forwarded to the RX, amplified by a CML-to-CMOS converter~\cite{7042350}, and a digitally-controlled delay line (DCDL)~\cite{8357042} is used to compensate for the skew between data and clock. 
\begin{figure}[!t]
\centering
\includegraphics[width=\columnwidth]{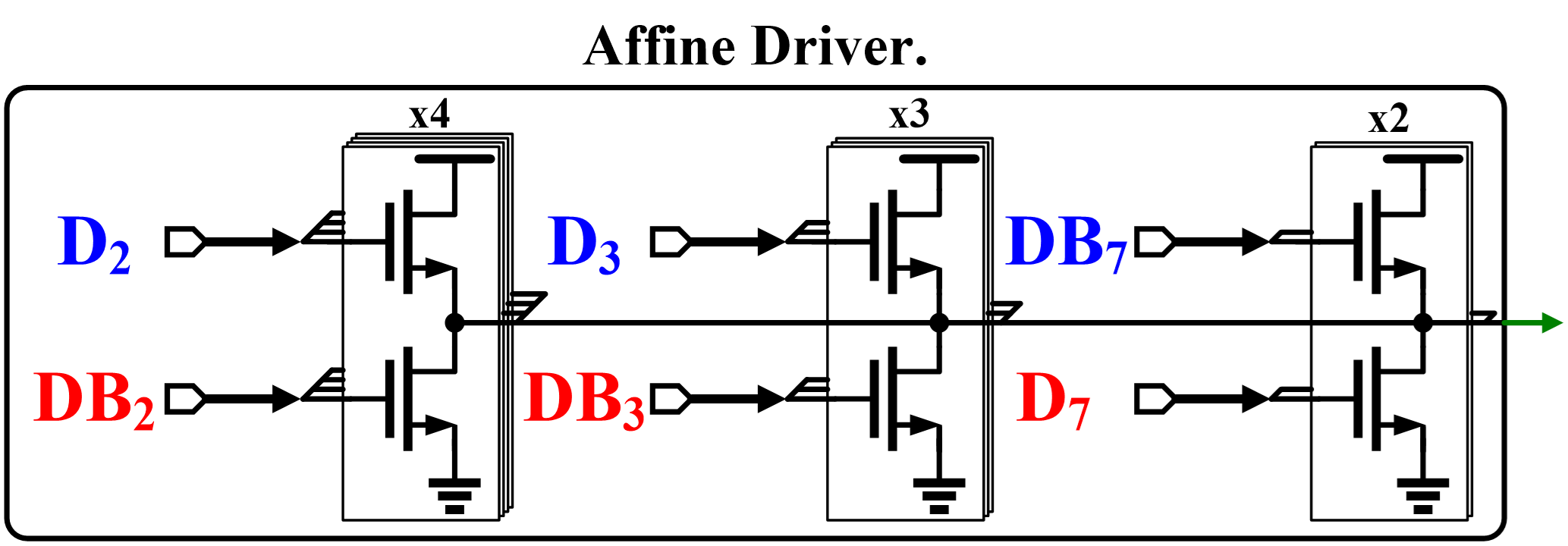}
\caption{Affine driver implementation corresponding to the row vector [0 4 3 0 0 0 -2].}
\label{fig:tx}
\end{figure}
\begin{figure}[!t]
\centering
\includegraphics[width=\columnwidth]{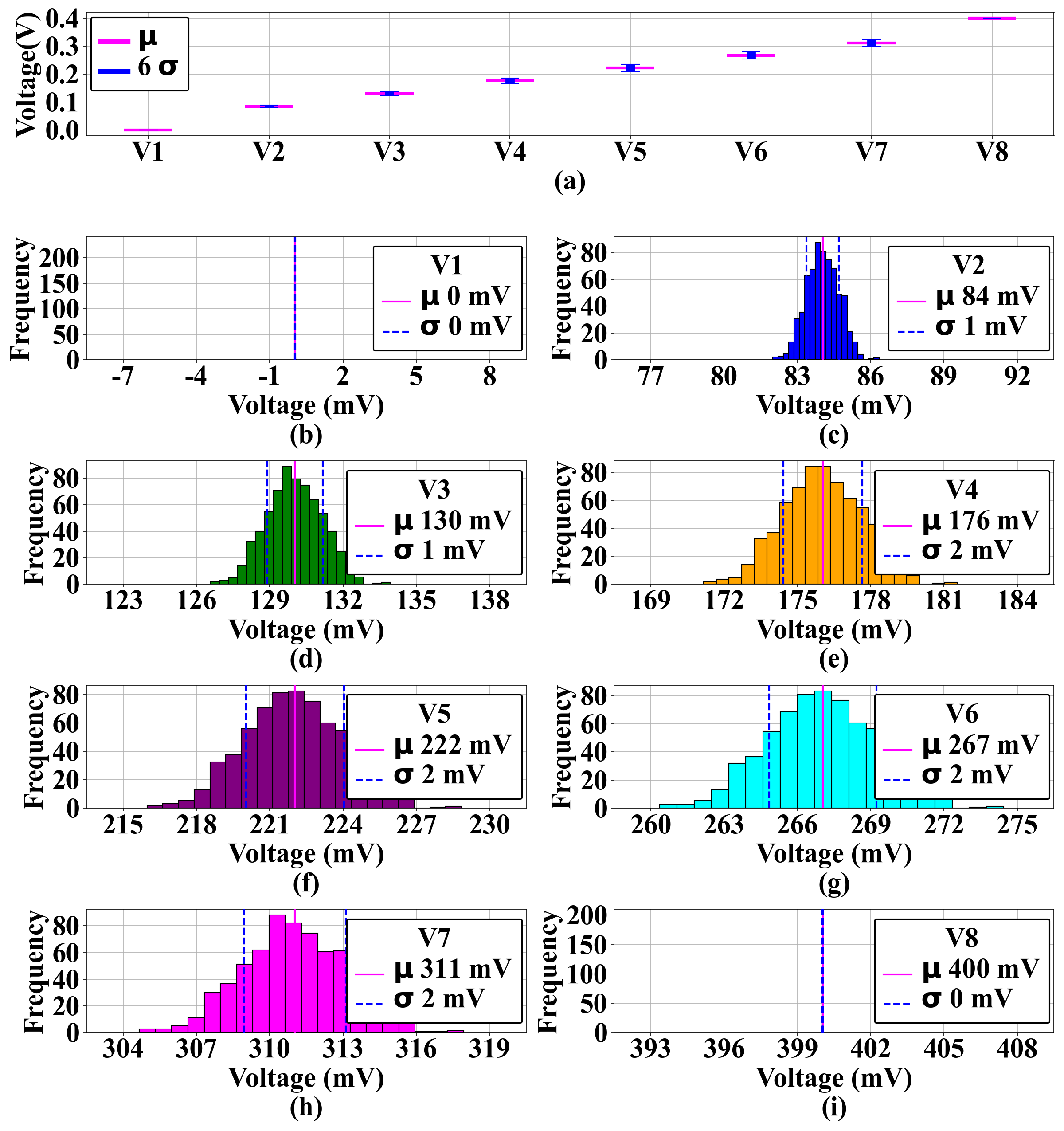}
\vspace{-1em}
\caption{Simulated affine driver output voltage variation.}
\label{fig:monte}
\end{figure}

The TX affine driver consists of multiple N-over-N drivers, where each row vector of $\mathbf{T}$ determines the weight of each driver.
For instance, as illustrated in Fig.~\ref{fig:tx}, the driver corresponding to the row vector [0 4 3 0 0 0 -2] consists of N-over-N drivers with weights of 4, 3, and 2, respectively. 
Pull-up inputs for each driver are $D_2$, $D_3$, and $\bar{D}_7$, and pull-down inputs are $\bar{D}_2$, $\bar{D}_3$, and $D_7$. 
This implementation generates the output voltage level of $V_{CM}+V_{R}(4D_2+3D_3-2D_7)$, which is an affine transformation of $(D_2, D_3, D_7)$. 
Supply voltage of the driver in the prototype is chosen to be 0.4\,V to minimize power.
Fig.~\ref{fig:monte} illustrates the device-mismatch-induced distribution of 8 analog voltage levels produced by the designed affine driver. 
It also presents the histograms of voltage distributions for each output level, 
which demonstrates that the output voltage level variation due to device mismatch is negligible.

\begin{figure}[!t]
\centering
\includegraphics[width=\columnwidth]{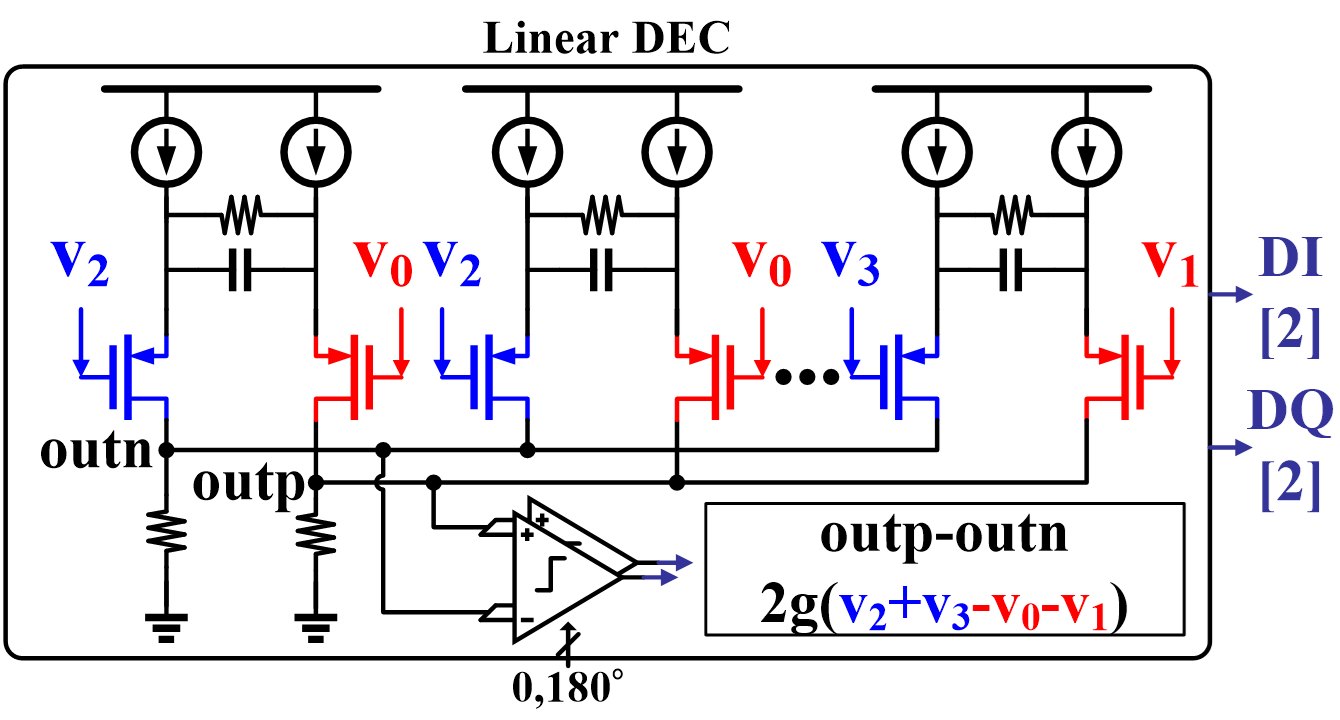}
\caption{Linear decoder implementation corresponding to the row vector [-2 -2 2 2 0 0 0 0].}
\label{fig:rx}
\end{figure}
Similar to the TX, the RX decoders at the front end are implemented, following the row vectors of $\mathbf{R}$, to restore the binary symbol through a linear transformation. 
For instance, Fig.~\ref{fig:rx} shows the decoder implementation corresponding to the row vector [-2 -2 2 2 0 0 0 0] in $\mathbf{R}$. 
Differential pairs with capacitive source degeneration allows RX to compensate for channel loss and to perform $v_3-v_1+v_4-v_2$ operation with some gain for recovering $D_3$. 
All the seven decoders employ the identical topology with the appropriately chosen channel inputs based on $\mathbf{R}$.

\begin{figure}[!t]
\centering
\includegraphics[width=\columnwidth]{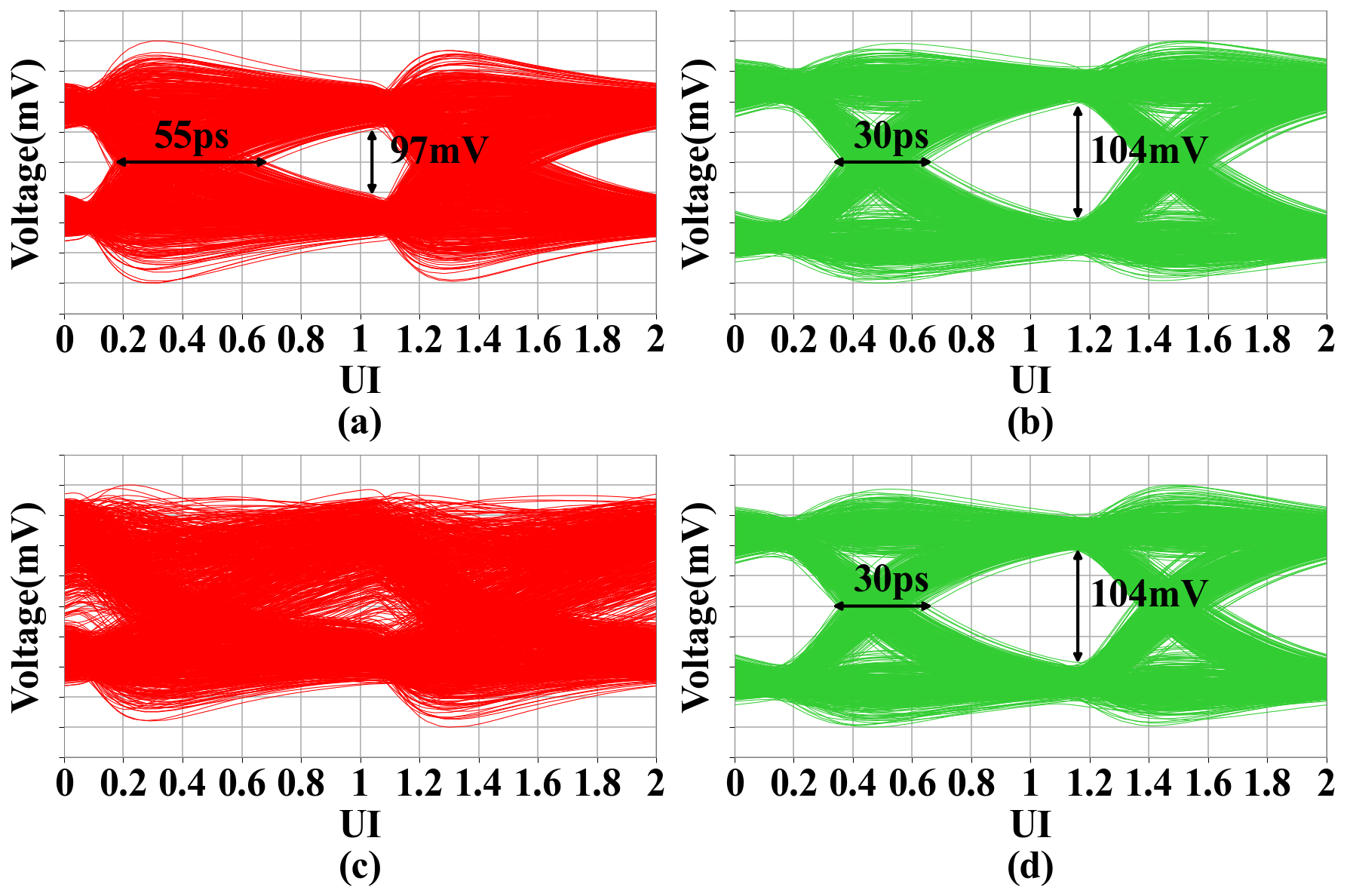}
\caption{RX eye diagrams: (a) SE without SSN, (b) XMAS without SSN, (c) SE with SSN, and (d) XMAS with SSN (5\,nH supply inductance).}
\label{fig:eyesim}
\end{figure}
\begin{figure}[!t]
\centering
\includegraphics[width=\columnwidth]{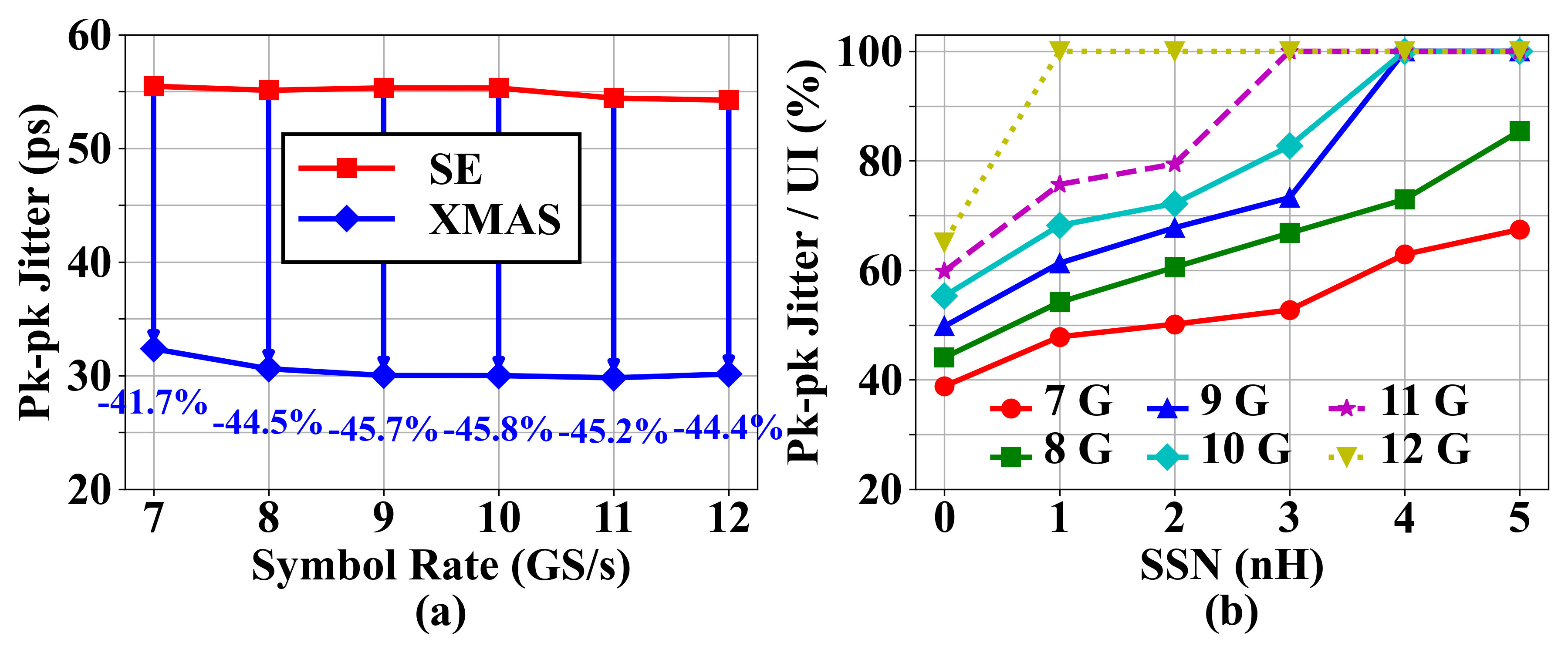}
\caption{(a) Peak-to-peak jitter reduction with XMAS across various symbol rates. (b) Peak-to-peak jitter with SE for various supply inductances and data rates. \ignore{(c) Peak-to-peak jitter with XMAS for various supply inductances and data rates.} }
\label{fig:cijcomp}
\end{figure}

We validate the effectiveness of the proposed XMAS interface through simulation results. 
As shown in Fig.~\ref{fig:eyesim}(a,b), compared to SE, thanks to its XTC performance XMAS reduces CIJ from 55\,ps to 30\,ps. 
This timing margin improvement becomes even more pronounced in the presence of SSN.
With a supply inductance of 5\,nH, as shown in Fig.~\ref{fig:eyesim}(c,d), while the SE eye completely closes, the XMAS eye is not degraded at all. 
In case of SE, even with small supply inductance, SE experiences huge peak-to-peak jitter caused by both SSN and CIJ, which leads to significantly degraded signal integrity especially at higher data rates (see Fig.~\ref{fig:cijcomp}(b)). 
On the other hand, XMAS does not suffer from SSN at all, and 
as shown in Fig.~\ref{fig:cijcomp}(a), XMAS consistently demonstrates around 45\,\% jitter reduction on average across various symbol rates compared to SE even in SSN-free environments. 

\begin{figure}[!t]
\centering
\includegraphics[width=\columnwidth]{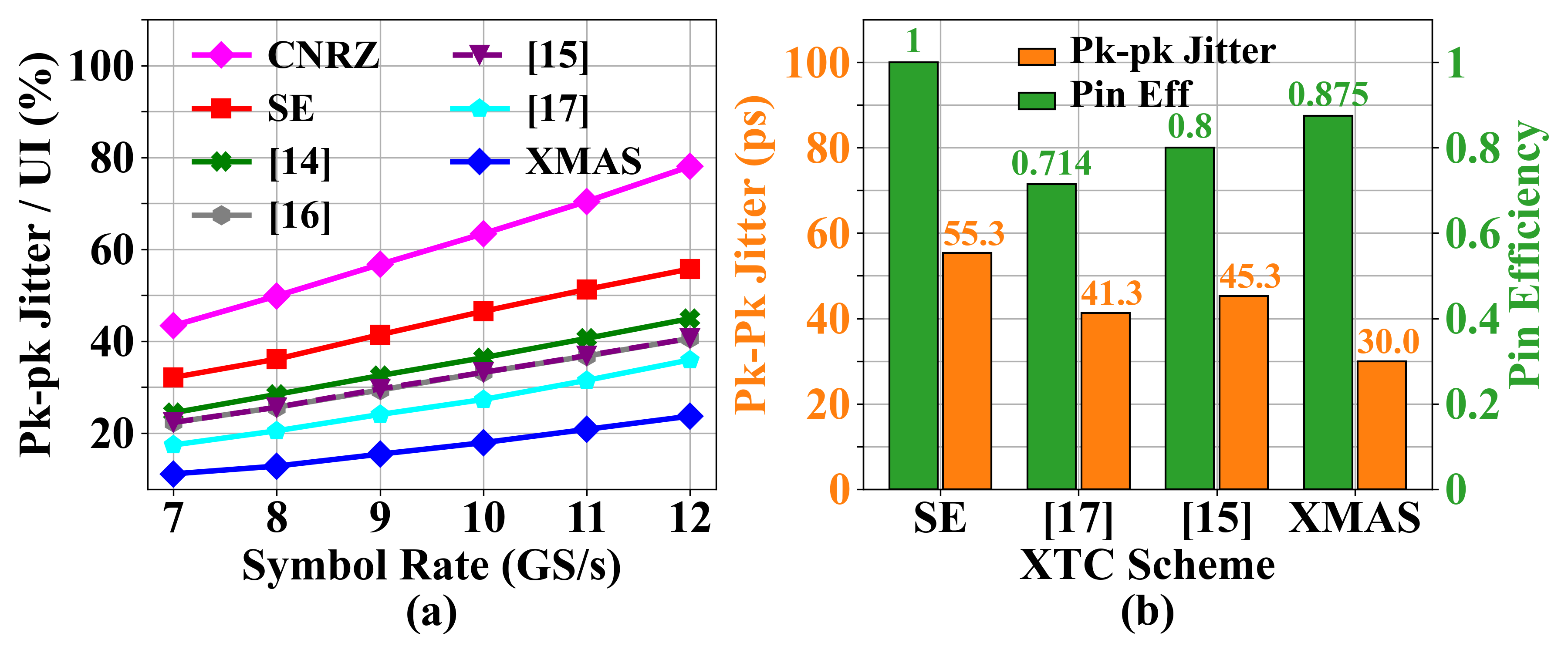}
\caption{(a) CIJ for various bus encoding techniques across different data rates. (b) Performance comparison of bus encoding techniques and XMAS. }
\label{fig:comparison}
\end{figure}
\begin{figure}[!t]
\centering
\includegraphics[width=\columnwidth]{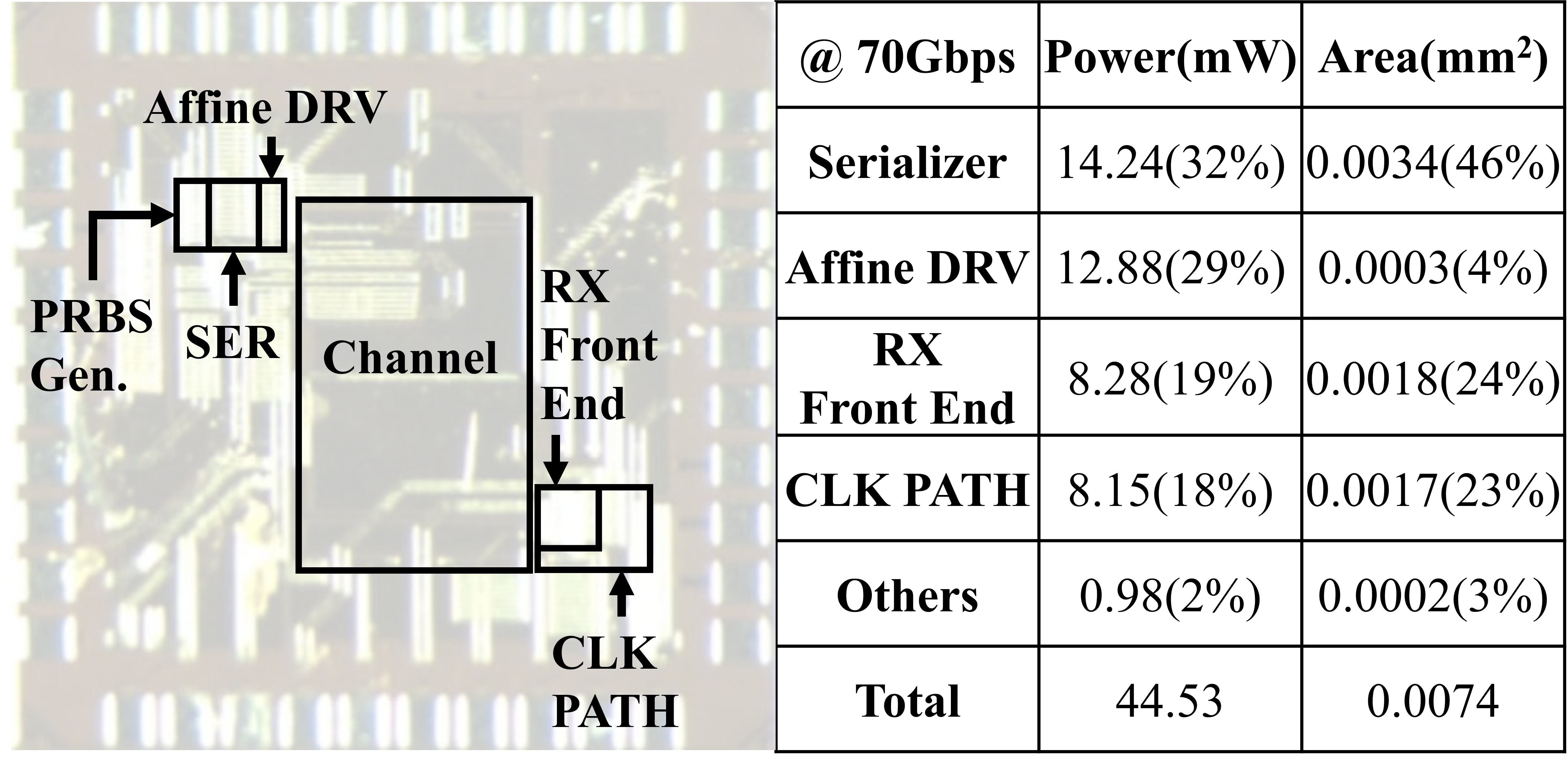}
\caption{Chip photomicrograph and power breakdown.}
\label{fig:photo}
\end{figure}


Prior bus coding techniques typically encode the data to avoid transmitting symbols sensitive to crosstalk by adding redundant bits, which limits the achievable pin efficiency. 
Comparison between prior art and XMAS is depicted in Fig.~\ref{fig:comparison}. 
For a fair comparison, all methods are compared at a fixed pin efficiency of 75\,\%\footnote{The XMAS matrices are modified to transmit 3 data over 4 channels for comparison.} with the identical channels. 
Among prior art, \cite{subrahmanya2004bus} achieves the highest XTC performance but uses a differential code, and \cite{liu20230} employs Fibonacci coding for XTC, where the encoders add logical distance between symbols.
These works use digital logic to perform encoding/decoding, which incurs a long end-to-end latency.  
On the other hand, since XMAS encodes/decodes the data without any digital logic, it does not incur any coding latency, while providing superior XTC performance. 
In the case of chord signaling~\cite{tajalli20201}, the XTC performance can vary significantly and may even boost the crosstalk depending on the MIMO channel characteristics. 
In other words, chord signaling~\cite{tajalli20201} does not necessarily guarantee XTC performance. 
Fig.~\ref{fig:comparison}(b) compares the pin efficiency and peak-to-peak jitter of XMAS with those of prior art with the highest XTC performance. 
In conclusion, XMAS shows the highest pin efficiency, while achieving the lowest CIJ without incurring additional coding latency. 


\ignore{
\subsection{Pseudo-Symmetric Channels}

To achieve the highest throughput, a specific physical structure shown in Fig.~\ref{fig:optimization}(a) is employed~\cite{ko20206}. 
The physical configuration of the channel determines characteristics related to crosstalk. 
An Impulse response of the channel, completely described by (2), result in a matrix illustrated in Fig. 4. 
Though the channel is not symmetrical due to adjacent channels' unequal crosstalk pulse response, XMAS can effectively handle it as if it were a symmetrical channel, given the presence of a linear relationship among the CPRs. 
}

\section{Measurement Results}
\label{sec:measured}

\begin{figure}[!t]
\centering
\includegraphics[width=\columnwidth]{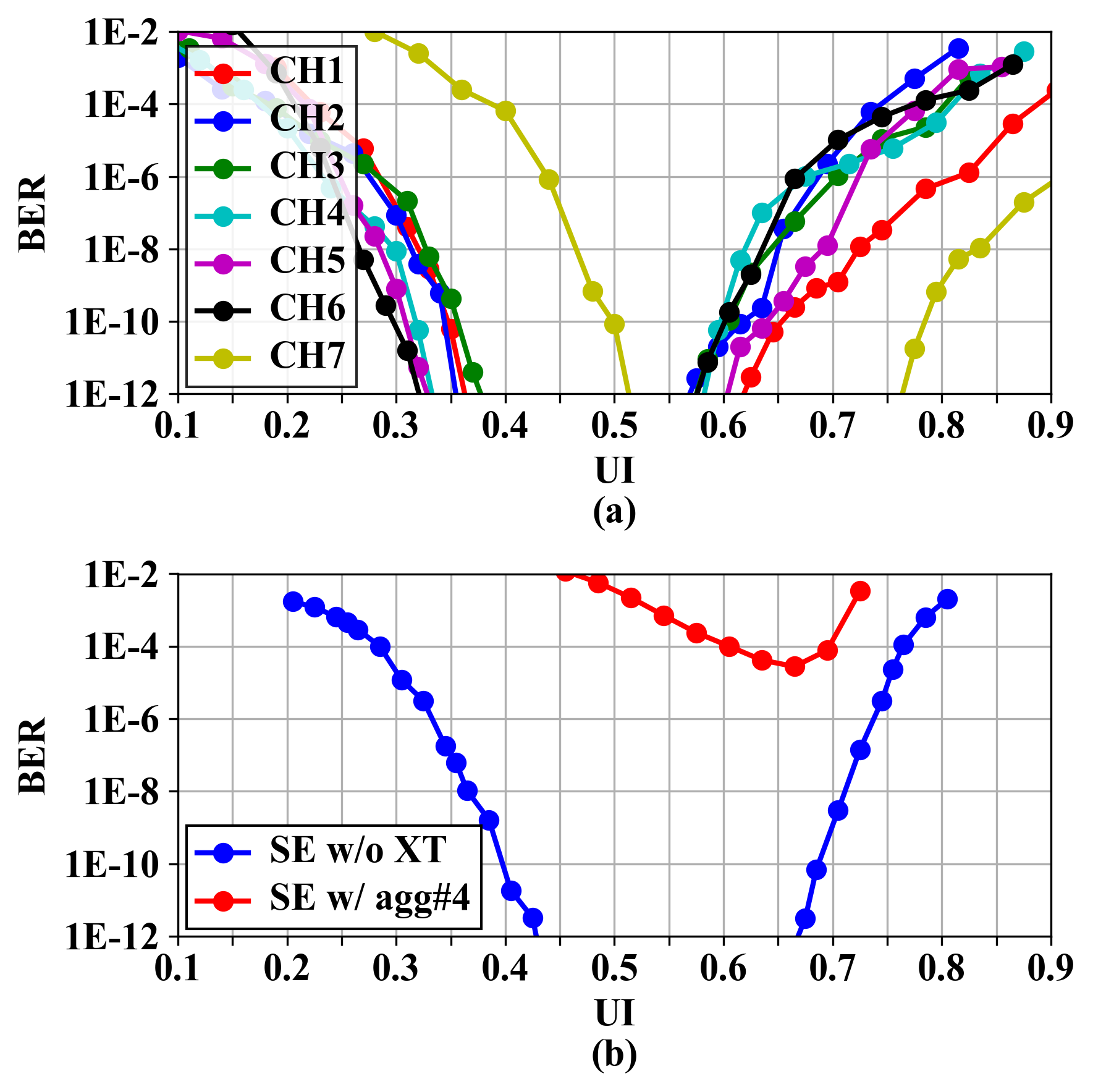}
\caption{Measured bathtub curves at 10\,Gb/s: (a) XMAS tested with PRBS15, and (b) SE tested with PRBS7.}
\label{fig:bathtub}
\end{figure}
\begin{figure*}[!t]
\centering
\includegraphics[width=\textwidth]{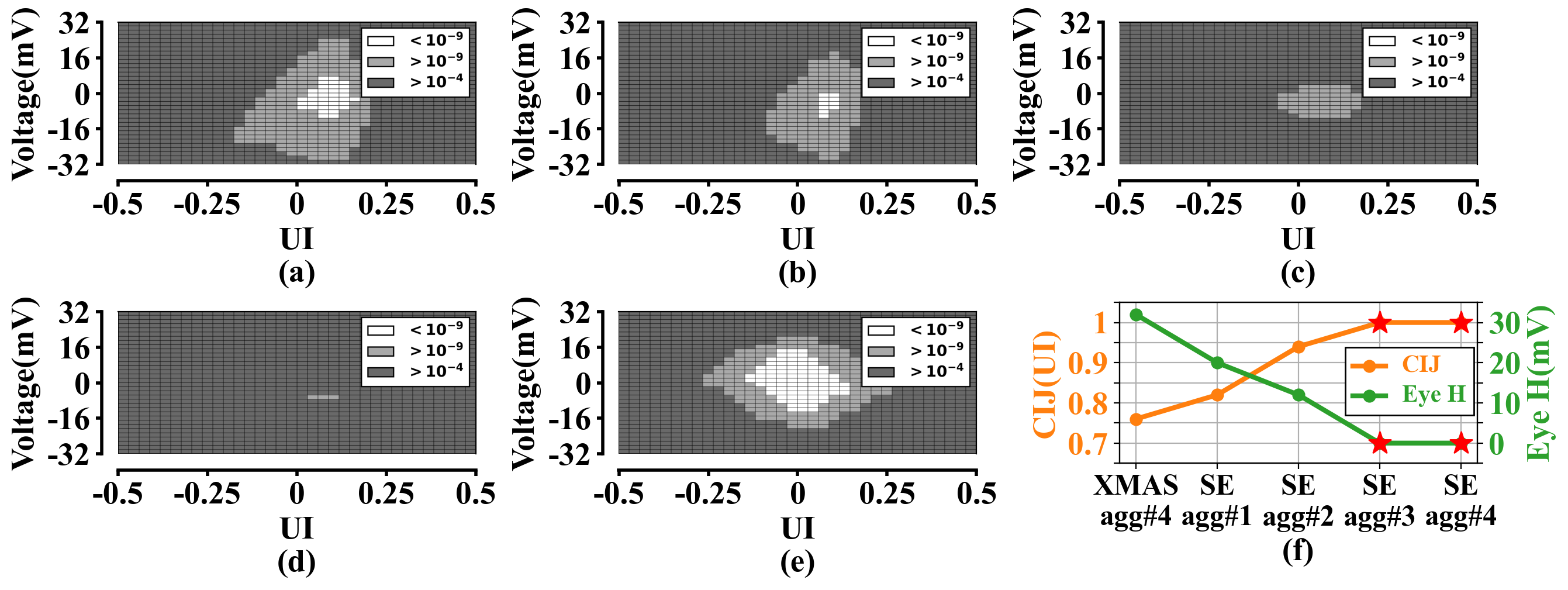}
\caption{Measured eye diagrams: (a) SE with 1 aggressor, (b) SE with 2 aggressors, (c) SE with 3 aggressors, (d) SE with 4 aggressors, (e) XMAS with 4 aggressors, and (f) eye height and peak-to-peak jitter across for different cases (BER at $10^{-9}$).}
\label{fig:eye}
\end{figure*}
The prototype transceiver was fabricated in a 28\,nm CMOS technology. 
Fig.~\ref{fig:photo} shows the die photo and the power breakdown of the chip. 
7-bit-parallel data are encoded and transmitted at 10\,GS/s/pin through eight channels (i.e., 70\,Gb/s aggregate bandwidth).
The overall transceiver occupies an active area of 0.0074\,mm$^2$ and consumes 44.53\,mW.
\ignore{, each 1.26\,mm long with a thickness of 0.36\,$\mu$m and a pitch of 0.126\,$\mu$m. 
The insertion loss of each channel at the Nyquist frequency is about -11\,dB, and the FEXT of the adjacent four channels is -24\,dB} 

\begin{figure}[!t]
\centering
\includegraphics[width=\columnwidth]{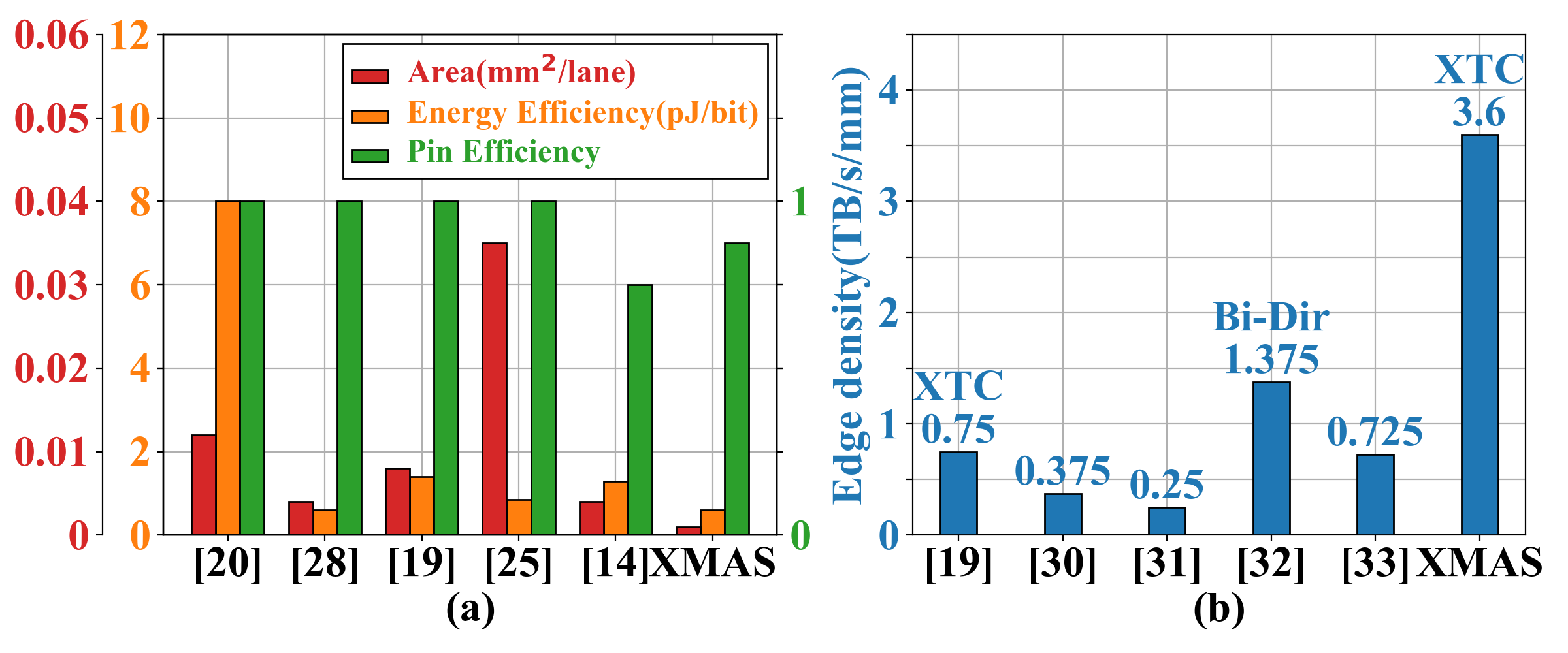}
\caption{Performance comparison: (a) area, power and pin efficiency with other XTC schemes, and (b) edge density with recent on-chip interfaces. }
\label{fig:papercomp}
\end{figure}
\ignore{
\begin{table}[!t]
\caption{\fixme{Measured crosstalk-induced jitter (CIJ)}\label{tab:table1}}
\vspace{-1em}
\centering
\includegraphics[width=3.5in]{CIJ_TABLE.png}
\vspace{-3em}
\end{table}
}

Fig.~\ref{fig:bathtub} shows the measured bathtub curves of XMAS and SE at 10\,Gb/s. 
As shown in Fig.~\ref{fig:bathtub}(a), the proposed XMAS allows all seven data to have a timing margin at least 0.2\,UI at a BER of 10$^{-12}$.
Since the test chip does not have any per-pin deskew circuit, $D_7$ has a timing skew of about 0.175\,UI, 
but the proposed XMAS achieves error-free operation for all the data.
However, the timing margin with SE is severely degraded with crosstalk, 
and BER lower than $10^{-5}$ cannot be achieved with four aggressors (see Fig.~\ref{fig:bathtub}(b)).
The voltage margin is also significantly improved with XMAS as demonstrated in Fig.~\ref{fig:eye}.
For SE, the eye width and height decrease rapidly as the number of aggressors increases. 
Even with three aggressors, SE cannot achieve BER less than 10$^{-9}$, while XMAS has a margin of 0.24\,UI width and 32\,mV height with four aggressors.
XMAS provides larger voltage and timing margin compared to SE with just one aggressor (see Fig.~\ref{fig:eye}(f)). Peak-to-peak jitter for SE without any crosstalk is measured to be 68\,ps, and Fig.~\ref{fig:eye}(f) shows that SE suffers from CIJ even with one aggressor, increasing jitter to 82\,ps, while XMAS is capable of greatly suppressing CIJ and jitter is only 76\,ps with four aggressors.

\ignore{
Table~\ref{tab:table1} shows how much crosstalk-induced jitter (CIJ) is reduced by XMAS. 
As the number of aggressors increases, jitter increases from 68\,ps to 100\,ps. 
Unlike SE, XMAS signaling combines eight wires, so reducing the number of aggressors below four is impossible. 
However, even comparing the worst case of Aggressor 4 with SE with one aggressor, a 42.8\% crosstalk-induced jitter reduction occurs. The larger the number of aggressors, which indicates the severity of XTALK, the greater the jitter reduction effect of XMAS. 
}

\begin{table*}[!t]

\caption{Comparison with state-of-the-art XTC schemes and on-chip interfaces.\label{tab:table2}}
\centering
\includegraphics[width=\textwidth]{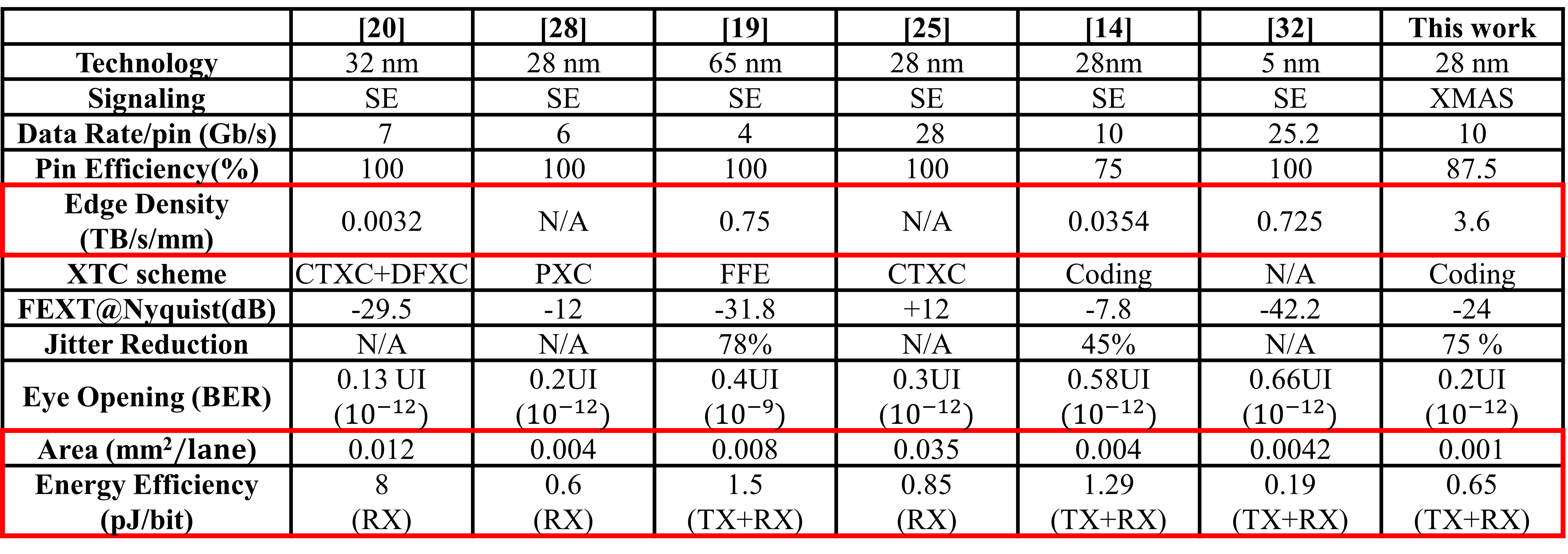}
\end{table*}

Compared with other state-of-the-art XTC schemes, XMAS shows the best XTC performance and achieves the highest edge density with comparable energy efficiency and area.
One popular XTC method~\cite{aprile2018eight} uses circuit-level techniques, where the receiver integrates a continuous-time compensation equalizer with a decision feedback equalizer (DFE). 
The DFE is further augmented with logic that removes post-taps due to crosstalk. 
However, this approach suffers from significant power and area overhead due to its complex equalizer configuration.
Another method~\cite{du2019compact} mitigates crosstalk by dividing the transmission bandwidth and splitting the phase in high-frequency bands. 
Though different from adjusting transmission delay for XTC~\cite{muljono20192}, it shares similarities with phase domain equalization as it separates noise sources in the phase domain and applies to the filter.
Coding-based XTC schemes, on the other hand, eliminates the need for complex equalizers and exhibit good area and power efficiency, but they generally face limitations such as low pin efficiency and large coding latency. 

Fig.~\ref{fig:papercomp}(a) illustrates the area, energy, and pin efficiency of various XTC schemes, where XMAS, despite being coding-based, shows almost the best energy efficiency and minimal hardware overhead with pin efficiency close to 1. 
Additionally, XMAS achieves the highest edge density compared to the studies over the past three years (see Fig.~\ref{fig:papercomp}(b))\cite{park202268,park20220,nishi20230,ko20206,nishi20231}).
Table~\ref{tab:table2} compares state-of-the-art interfaces over the past five years. 
Notably, the XMAS transceiver excels by achieving a great edge density of 3.6\,TB/s/mm, an energy efficiency of 0.65\,pJ/bit, and an area efficiency of 0.0012\,mm\text{$^2$}/lane. 

\section{Conclusion}
\label{sec:conclusion}
This paper introduces an I/O interface with the novel signaling XMAS, which is designed to support high-speed data transmission over channels with high crosstalk. 
XMAS not only demonstrates exceptional crosstalk removing capabilities but also exhibits robustness against noise, especially simultaneous switching noise. 
To maximize the edge density, co-optimizing the design of channels and signaling is performed using an analytical model of XMAS. 
The prototype XMAS transceiver is fabricated in a 28-nm CMOS process and achieves an edge density of 3.6\,TB/s/mm with an energy efficiency of 0.65\,pJ/b. 
Compared to SE, CIJ of the received eye with XMAS is reduced by 75\,\% at 10\,GS/s/pin data rate, and the horizontal eye opening extends to 0.2\,UI at a bit error rate less than 10\text{$^{-12}$}.

\section*{Acknowledgments}
The EDA tool was supported by the IC Design Education Center (IDEC), Korea.

\bibliography{ref}
\bibliographystyle{IEEEtran}

\vfill

\end{document}